\documentclass{revtex4}    
\usepackage{amssymb,epsf}
\usepackage{latexsym}
\begin{document}
\title{Exact String-like Solutions of the Gauged Nonlinear O(3) Model}
\author{Hamid Reza \surname{Vanaie}}
\author{Nematollah \surname{Riazi}}
\address{Physics Department,
\\Shiraz University,
Shiraz 71454, Iran,\\
and\\
IPM, Farmanieh, Tehran, Iran.\\
email: {\rm riazi@physics.susc.ac.ir}}

%
\begin{abstract}
We show that the least energy conditions in the gauged nonlinear
sigma model with Chern-Simons term lead to exact soliton-like
solutions which have the same features as domain walls. We will
derive and discuss the corresponding solutions, and compute the
total energy, charge, and spin of the resulting system.
\end{abstract} \maketitle

\section{Introduction}

The O(3) sigma model in (2+1) dimensions is a popular model in
theoretical physics and has been studied extensively
\cite{sch}\cite{bel}. As far as static solutions are concerened,
the model is integrable and of Bogomol'nyi type \cite{bog}. The
exact solutions are therefore available via simple analytical
expressions. In contrast, similar models in (3+1) dimensions
(\cite{bar},\cite{pey},\cite{pie},\cite{sky}), are neither
integrable nor of Bogomol'nyi type. Nevertheless, the soliton
solutions of models in 3+1 dimensions are extensively studied
numerically and have found application in particle and nuclear
physics \cite{wit}. Soliton solutions in (2+1) dimensions,
modified by the addition of the Hopf term are topologically stable
and  are classified according to the homotopy $\pi_2(S^2)=Z$.
These solutions are scale-invariant. This property leads to some
difficulties upon quantization. Moreover, upon interacting with
each other, solitons undergo change of size and tend to become
singular \cite{lee}. The soliton solutions of the O(3) sigma model
with a suitable choice of potential are presented explicitly in
\cite{lee2} together with their interaction behavior. These
solitons are also called Q-lumps.

There are two natural ways for breaking the scale-invariance:
gauging the symmetry, and  including a potential term \cite{sch}.
It has been shown that gauging the U(1) subgroup by a Chern-Simons
term and including  a suitable potential leads to both topological
and nontopological solitons which are not scale-invariant
\cite{gho}. The potential discussed in \cite{gho} has two discrete
minima $\phi_3=\pm 1$ for which the U(1) symmetry is not
spontaneously broken in (2+1) dimensions. This model was further
studied by Mukherjee \cite{muk}, where it was shown that the
degeneracy of topological solitons is removed by including a
potential with symmetry-breaking minima. Solutions with azimuthal
vector potentials (${\bf A}=\hat{\theta}\frac{n}{r}$), have
quantized magnetic flux ($\Phi=2\pi n$), spin ($S=-\pi kn^2$), and
charge ($Q=-2\pi kn$).

In the present paper, we will show that within the framework of
the model presented in \cite{muk}, exact soliton solutions exist
which depend on one spatial dimension, resembling domain-wall
solutions of the $\phi^4$ theory. Because of the non-trivial
mapping between the boundaries and the discrete vacua, the
solutions are topologically stable. The stability and lack of the
scale invariance are also verified using the energy functional of
the system.

\section{Gauged O(3) sigma model}

  The lagrangian of the  model considered here, is the one partially studied in
\cite{muk}, in (2+1) dimensions:
\begin{equation} {\cal L}=\frac{1}{2}D_\mu \phi.D^\mu
\phi+\frac{k}{4}\epsilon^{\mu\nu\lambda}A_\mu \partial_\nu
A_\lambda -U(\phi). \end{equation}
 Here  $\phi=(\phi_1,\phi_2,\phi_3)$ is a triplet of scalar
 fields, constrained to a $S^2$:
 \begin{equation}
 \phi_a\phi_a =1,
 \end{equation}
 where
\begin{equation} \phi_a=n_a.\phi , \end{equation} and $n_a$ (a=1,2,3) are unit
orthogonal vectors in the internal space. We work in the Minkowski
spacetime with the metric tensor $g_{\mu\nu}=(1,-1,-1)$.
 $D_\mu \phi$ is the covariant derivative given by \cite{sch}:
 \begin{equation}
 D_\mu \phi=\partial_\mu \phi +A_\mu {\bf n}\times \phi ,
 \end{equation}
 where ${\bf n}$ is the (constant) unit vector defined by
 \begin{equation}
{\bf n}=\lim_{|x|\rightarrow \infty} {\bf \phi}(t,{\bf x}).
\end{equation}

The U(1) subgroup is gauged by the vector potential $A_\mu$ whose
dynamics is dictated by the Chern-Simons term. The potential
\begin{equation} U(\phi )=\frac{1}{2k^2}\phi_3^2(1-\phi_3^2) \end{equation}
gives a self interaction of the field $\phi_a$. Note that the
minima of the potential reside either on,
\begin{equation}\label{vac1} \phi_1=\phi_2=0\ \ {\rm and}\ \
\phi_3\pm 1, \end{equation}
                 or
\begin{equation}\label{vac2} \phi_3=0\ \ {\rm and}\ \ \phi_1^2+\phi_2^2=1. \end{equation} In 2+1
dimensions, soliton solutions of  this system has been studied
\cite{muk}. In such a case, the U(1) symmetry  is unbroken in
(\ref{vac1}), whereas (\ref{vac2}) corresponds to the spontaneous
breaking of this symmetry.

\begin{figure}[t]
 \epsfxsize=10cm
  \centerline{\epsffile{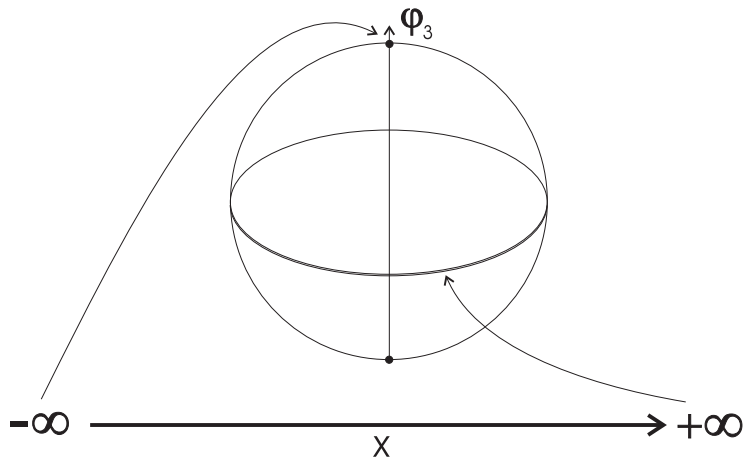}}
  \caption{String-like solutions correspond to mappings between the boundaries of $x$ and
  the disconnected vacua of the $\phi$-fields.}
  \label{vacua}
\end{figure}

In 1+1 dimensions, we expect that the disconnected vacua
facilitate the existence of solitons. Because of the existence of
disconnected  vacua, we can have non-trivial mappings between the
boundaries of the 1-dimensional space and these disconnected vacua
and thus topological stability (Figure \ref{vacua}). In order to
obtain these soliton solutions, we start with Schwinger's
energy-momentum tensor \cite{shw}, which in the static limit
becomes:
\begin{equation}\label{energeq} E=\frac{1}{2}\int d^2x \left( D_i \phi . D_i \phi
+\frac{k^2B^2}{1-\phi_3^2}+\frac{1}{k^2}\phi_3^2(1-\phi_3^2)\right),
\end{equation} where $B=curl(A)$ is the magnetic field.
A conserved current can also be constructed: \begin{equation}
K_\mu =\frac{1}{8\pi}\epsilon_{\mu\nu\lambda}\left[ \phi .D^\nu
\phi \times D^\lambda \phi -F^{\nu\lambda}\phi_3\right].
\end{equation} For static solutions which depend on one space coordinate
(say $x$), $K_o$ becomes:
\begin{equation}\label{topcharge} K_o=-\frac{1}{4\pi}\left[ A_2\phi_3\frac{\partial
\phi}{\partial x}.\phi +\frac{\partial \phi_3}{\partial x}\right]
+\frac{1}{4}\phi_3\frac{\partial A_2}{\partial x}, \end{equation}
and the corresponding conserved total charge is: \begin{equation}
T=\int dx K_o.
\end{equation}
We also have
\begin{equation}
B=F_{12}=\frac{\partial A_2}{\partial x}, \end{equation} with
\begin{equation}
A_2=\frac{\phi_3}{k},
\end{equation}
which leads to
\begin{equation}
B=\frac{1}{k}\frac{d\phi_3}{dx}.
\end{equation}

Rearranging (\ref{energeq}), we can write \cite{gho}\cite{raj}:
\begin{equation} \label{energy} E=\int
d^2x\left[ \frac{1}{2}(D_i\phi\pm \epsilon_{ij}\phi\times D_j\phi
)^2+\frac{k^2}{1-\phi_3^2}\left( F_{12}\pm
\frac{1}{k^2}\phi_3(1-\phi_3^2)\right)^2\right] \pm 4\pi T.
\end{equation}
This leads to the Bogomol'nyi conditions:
 \begin{equation}
 \label{bog1}
 D_i\phi\pm \epsilon_{ij}\phi\times D_j\phi=0,
 \end{equation}
 and
 \begin{equation}
 \label{bog2}
 F_{12}\pm \frac{1}{k^2}\phi_3(1-\phi_3^2)=0,
 \end{equation}
which minimize the energy functional in a particular topological
sector.

\section{Exact Domain wall solutions}

\begin{figure}[t]
 \epsfxsize=10cm
  \centerline{\epsffile{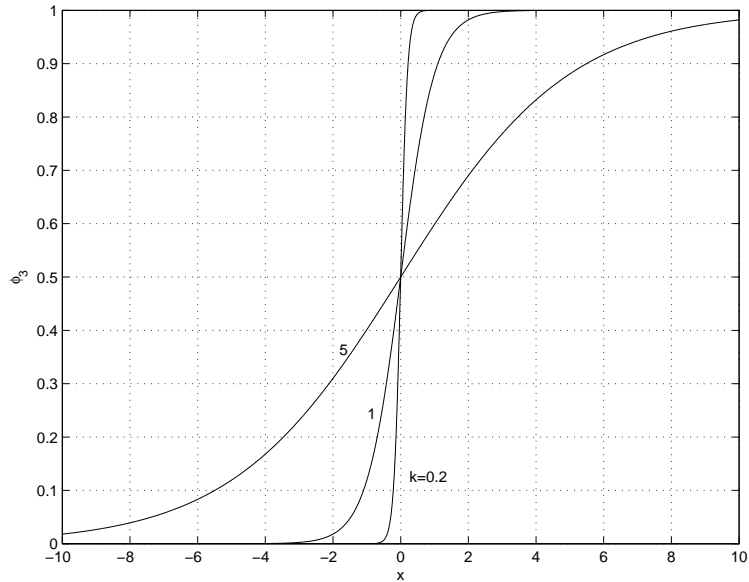}}
  \caption{Variation of the $\phi_3$ field as a function of the $x$-coordinate for $k=0.2,1.0,\ {\rm and}\ 5$.}
  \label{fig1}
\end{figure}

\begin{figure}[t]
 \epsfxsize=10cm
  \centerline{\epsffile{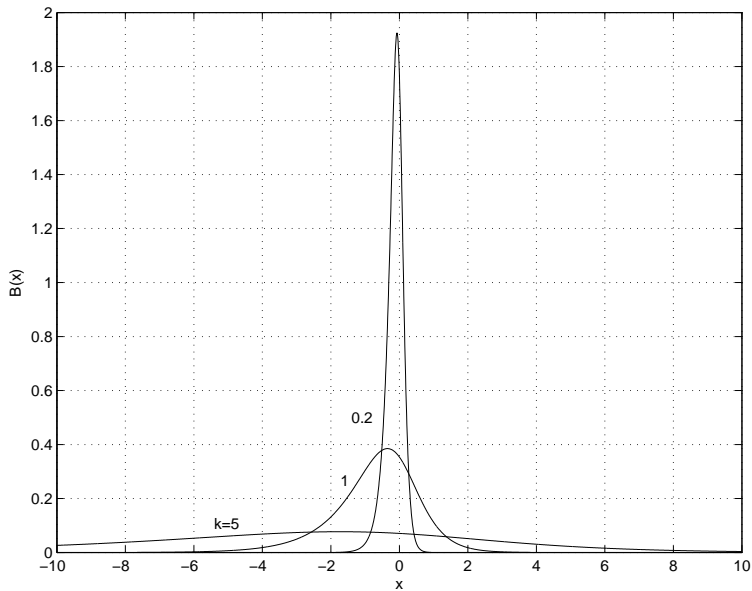}}
  \caption{Variation of the magnetic field $B$ as a function of the $x$-coordinate for $k=0.2,1.0,\ {\rm and}\ 5$.}
  \label{fig2}
\end{figure}

In this section, we obtain analytical domain-wall-like solutions
in the case where the fields depend on only one of the space
coordinates (say $x$). The $\phi$-fields, which are constrained on
a $S^2$, are assumed to vary on a longitude. Because of the U(1)
subgroup, this longitude is arbitrarily rotated to $\phi_2=0$, in
order to simplify the solutions. Let us expand the Bogomol'nyi
equations (\ref{bog1}) and (\ref{bog2}) as: \begin{equation}
\frac{\partial \phi_1}{\partial x}\mp \phi_3\phi_1A_2=0,
\end{equation} \begin{equation} \frac{\partial \phi_3}{\partial x}\pm A_2\mp
\phi_3^2A_2=0, \end{equation} \begin{equation} \pm
A_1\phi_3\phi_1=0, \end{equation}
\begin{equation}
\phi_1A_2\pm (\phi_1\frac{\partial \phi_3}{\partial
x}-\phi_3\frac{\partial \phi_1}{\partial x})=0, \end{equation} and
\begin{equation} \mp A_1\pm A_1\phi_3^2 =0. \end{equation}
It can be easily shown that the above equations, together with
appropriate boundary conditions, lead to:
\begin{equation}\label{dphi3} \frac{\partial \phi_3}{\partial
x}-A_2(1-\phi_3^2)=0,
\end{equation} and \begin{equation} A_1(1-\phi_3^2)^{1/2}=0, \end{equation}
which implies \begin{equation} A_1=0.\end{equation}     Note that
$\phi_3^2\neq 1\ \ \forall x$ for the desired solutions. We also
have
\begin{equation}\label{da2} \frac{\partial A_2}{\partial x}\pm
\frac{1}{k^2}\phi_3(1-\phi_3^2)=0.\end{equation} Equations
(\ref{dphi3}) and (\ref{da2}) are readily solved to obtain
\begin{equation}
\phi_3=\pm \frac{1}{(1+e^{-\frac{2x}{k}})^{1/2}},
\end{equation}
\begin{equation} A_1=0.\end{equation}and\begin{equation}
A_2=\frac{\phi_3}{k},
\end{equation} From these, the $B$ field is also calculated:
\begin{equation} B=F_{12}=\frac{\partial A_2}{\partial x}, \end{equation} or
\begin{equation} B=\pm \frac{1}{k^2}\frac{\partial \phi_3}{\partial
x}=\frac{e^{-\frac{2x}{k}}}{k(1+e^{-\frac{2x}{k}})^{3/2}}.
\end{equation}
$\phi_3(x)=kA_2(x)$ is plotted in Figure \ref{fig1}, and $B(x)$ is
shown in Figure \ref{fig2}.
\section{Conserved Quantities and Stability}
\begin{figure}[t]
 \epsfxsize=10cm
  \centerline{\epsffile{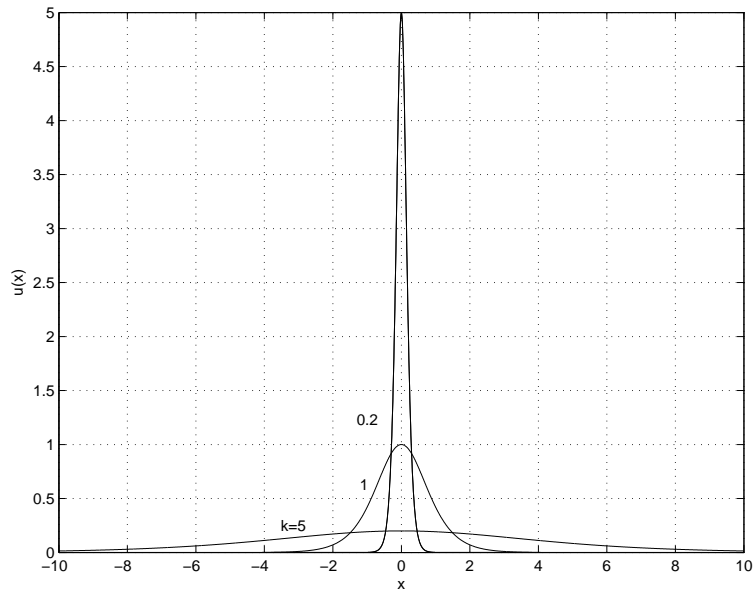}}
  \caption{Variation of the energy density $u(x)$ as a function of the $x$-coordinate for $k=0.2,\ 1.0,\ {\rm and}\ 5.0$.}
  \label{fig3}
\end{figure}

Using (\ref{energeq}), the energy density can be easily calculated
\begin{equation} u=\frac{4}{k^2}\frac{e^{-2x/k}+e^{-4x/k}}{(1+e^{-2x/k})^3}, \end{equation}
which is plotted in Figure (\ref{fig3}). It is seen that the
magnetic field and  energy are  concentrated in a thin region
around $x=0$, quite similar to the domain wall solutions of the
$\phi^4$ theory in (3+1) dimensions. Here, the solution
corresponds to a string on the $xy$-plane which extends along the
$y$-axis. Integrating the energy density over $x$, the total
energy per unit length of the string is obtained:
\begin{equation} E=\int_{-\infty}^{+\infty}u(x)dx=\frac{2}{k^2}. \end{equation} In order to check for the stability of
the solution, we apply the scale transformation \begin{equation}
x\rightarrow \alpha x \end{equation} which leads to
\begin{equation} E\rightarrow E(\alpha)=\frac{1}{\alpha}f+\alpha
g,\end{equation} where $f$ and $g$ are positive quantities
independent of $\alpha$. The energy is found to be a minimum at
$\alpha =1$, signalling the stability.

\begin{figure}[t]
 \epsfxsize=10cm
  \centerline{\epsffile{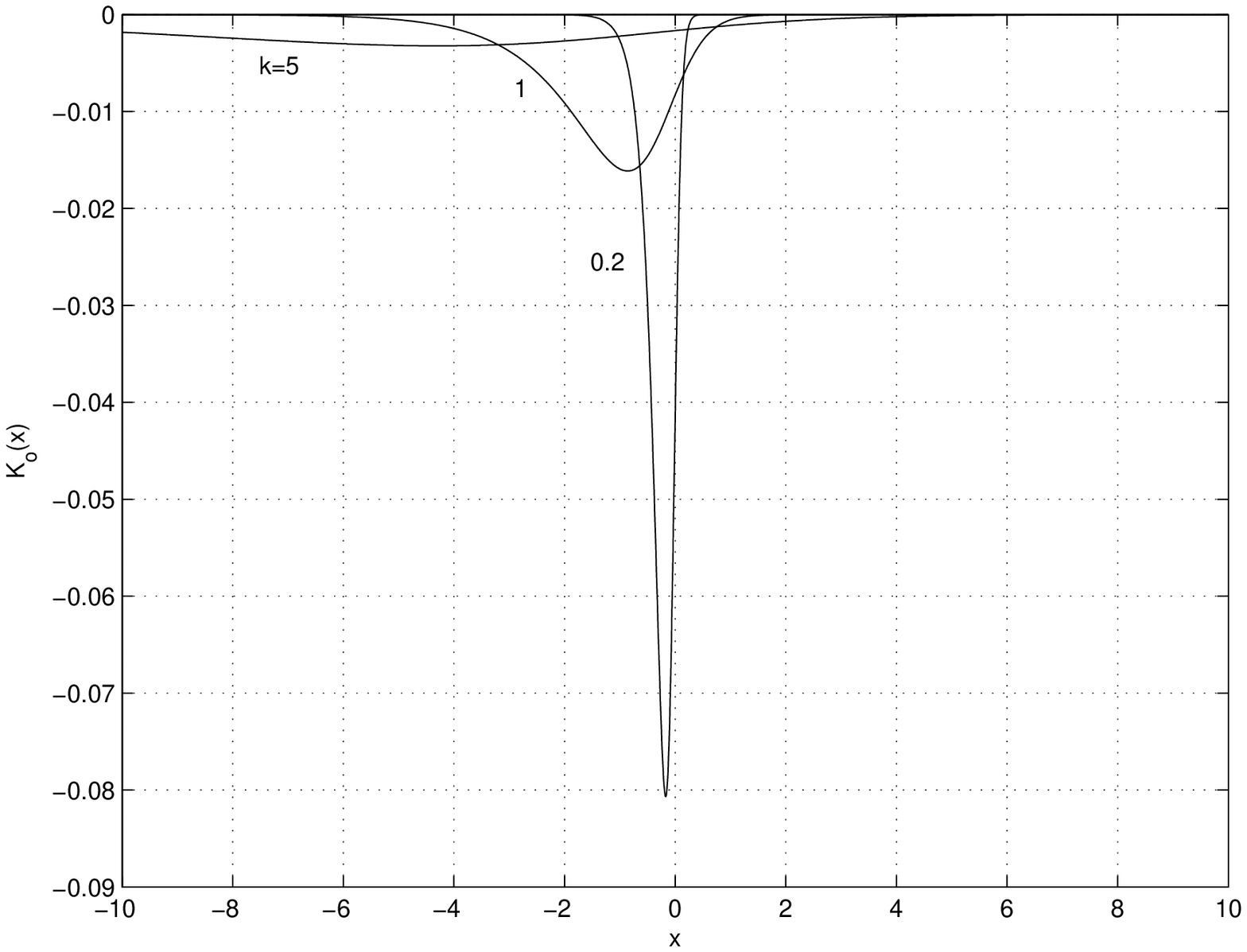}}
  \caption{Variation of the charge density $K_o$ as a function of the $x$-coordinate for $k=0.2,\ 1.0,\ {\rm and}\ 5.0$.}
  \label{fig4}
\end{figure}

Let us calculate the topological charge density (the zero
component of topological current). From (\ref{topcharge}), we
obtain \begin{equation} K_o=\frac{1}{4\pi
k^2}\frac{e^{-2x/k}}{(1+e^{-2x/k})^{3/2}}\left[
\frac{1}{(1+e^{-2x/k})^{1/2}}-1\right],
\end{equation} which is shown in Figure (\ref{fig4}). The total
topological charge is subsequently obtained: \begin{equation}
T=\int_{-\infty}^{+\infty}K_o(x)dx=-\frac{0.0398}{k^2}\ \ \ {\rm
(numerical)}.
\end{equation}
The values of $E$ and $T$ are seen to satisfy the well known
inequality $E\geq 8\pi T$.

The spin of the solution \cite{ban1a}\cite{ban2} is easily shown
to vanish:
\begin{equation}
S_i=-\frac{k}{2}\oint_{boundary}(x_iA^2-A_ix_jA^j)n^idl=0.\end{equation}

\section{Conclusion}

In a field theory possessing the vacuum manifold ${\cal M}$,
formulated in a D-dimensional space plus 1 time dimension, the
localized static solutions constitute a mapping $\pi:\partial
M\rightarrow \cal{M}$, where $\partial M$ is the boundary of the
D-dimensional space. If the homotopy group of this mapping is
nontrivial, then topological, soliton-like solutions are expected
\cite{ria}. We showed that the nonlinear O(3) sigma model with a
Chern-Simons term and a suitable potential is such a model. We
obtained examples of these solutions which satisfy the
Bogomol'nyi's conditions. On the $xy$-plane, solutions correspond
to a string along the $y$-axis, with energy density, charge
density, and magnetic field confined to a thin band near to $x=0$.
The width of this band (string) depends on the value of the
parameter $k$. We also showed that the solutions are not scale
invariant under $x\rightarrow \alpha x$. Rather, the energy is
minimum for $k=1$, indicating the stability of the solutions.

\acknowledgements N. Riazi acknowledges the support of Shiraz
University and IPM.


\end{document}